# Ultrafast Electron Dynamics in Thiolate-Protected Plasmonic Gold Clusters: Size and Ligand Effect

[a]Masoud Shabaninezhad, [b]Abubkr Abuhagr, [c]Naga Arjun Sakthivel, [c]Chanaka Kumara, [c]Amala Dass*, [d]Kyuju Kwak, [d]Kyunglim Pyo, [d]Dongil Lee* and [b]Guda Ramakrishna*

[a]Department of Physics, Western Michigan University, Kalamazoo, MI 49008, United States

[b]Department of Chemistry, Western Michigan University, Kalamazoo, MI 49008, United States

[c]Department of Chemistry and Biochemistry, University of Mississippi, Oxford, MS 38677, United States

[d]Department of Chemistry, Yonsei University, Seoul 03722, South Korea

**Abstract**

The influence of passivating ligand on electron-phonon relaxation dynamics of the smallest sized gold clusters was studied using ultrafast transient absorption spectroscopy and theoretical modeling. The electron dynamics in $Au_{279}$, $Au_{329}$, and $Au_{329}$ passivated with 4-tert-butylbenzene thiol (TBBT), phenylethane thiol ($SC_2Ph$) and hexane thiol ($SC_6$), respectively, were investigated. These clusters were chosen as they are the smallest gold clusters reported till-date to show plasmonic behavior. Ultrafast transient absorption measurements were also carried out on $Au_{\sim1400}$ ($SC_6$) and $Au_{\sim2000}$ ($SC_6$) to understand the influence of the size on electron-phonon relaxation with the same passivating ligand. The study has revealed interesting aspects on the role of ligand on electron-phonon relaxation dynamics wherein the aromatic passivating ligands, SC$_2$Ph and TBBT, have shown smaller power dependence and higher plasmon bleach indicating dampened plasmon resonance while the cluster with aliphatic passivating ligand has behaved similarly to regular plasmonic gold nanoparticles. To model the effect of the ligand on the plasmonic properties of the investigated samples , free electron density correction factor of each one was calculated using three-layered Mie theory, and the results show that SC$_6$ interacts least with core-gold while TBBT and SC$_2$Ph have a greater effect on the surface electronic conductivity that is attributed to π-interaction of the ligand with gold. The results also shed light on unusual electron-phonon relaxation and smaller slope observed for Au$_{329}$ (SC$_2$Ph) that is ascribed to surface gold-π interaction creating a hybrid state. In contrast, extended π-interaction is probably the reason for plasmonic nature observed in Au$_{279}$ (TBBT) even though its size is smaller when compared to Au$_{329}$. In addition, the results also have shown that the electron-phonon coupling has increased with an increase in the size of the cluster and theoretical modeling has shown higher electron conductivity for larger plasmonic gold clusters.

## 1. Introduction

Optical properties of noble metal nanoparticles have been the subject of widespread research interest for the past two decades.[1–4] The coupling of the electromagnetic field to free electrons in metal nanoparticles creates quasi-particles named surface plasmons and was the focus of research for both theoreticians and experimentalists alike.[5] The surface plasmon resonance (SPR) of metal nanoparticles (NPs) strongly depend on the density of electrons, shape, type, size, the composition of the nanoparticles, the polarization direction of the incident light as well as the chemical environment that surrounds them.[5–16] Among metallic nanoparticles, gold NPs have attracted significant interest owing to their high chemical and physical stability, biocompatibility, photo-stability and large optical cross sections.[17–19] The SPR of gold NPs found applications in manifold or areas[20] that included optical sensing,[21,22] biolog-ical imaging,[23–25] plasmonic photo-thermal therapy,[4] molecular diagnostics,[2,24,26] surface-enhanced Raman spectroscopy,[2] metal-enhanced luminescence and plasmonic rulers.[3]

Enormous research has focused on the plasmonic properties of Au NPs and significant breakthroughs were made in the field.[2,3,11,14,17,19,24–31] However, a bulk of the surface plasmon research has focused on gold nanoparticles passivated with surfactants or gold nanomaterials in different matrices.[27,32–38] The SPR properties seem to be influenced by the chemical environment around the NPs and can be greatly altered if the surface of the gold nanoparticles is chemically bound to ligands.[27,32–35] The ligands that are chemically bound to Au NPs prevent coalescence of the NPs by counterbalancing the van der Waals attractive forces.[39–41] Coupling NPs with chemical ligands potentially reduce the electron density of the conduction electrons and thereby altering the effective optical refractive index at the near field of the NPs.[40–42] A remarkable work was carried out by Peng et. al[41] on the plasmon absorption spectrum of spherical silver (Ag) NPs with diameters in the range of 2 to 20 nm. They observed an interesting trend where the absorption peak of the ligand-conjugated Ag NPs shifted to higher energies when the size was decreased from 20 nm to 12 nm, while the absorption peak shifted to longer wavelengths with the further reduction of the size. This unusual trend was assigned to the ligand effect on SPR absorption.

Although significant progress was made on gold nanoparticles of varying sizes and shapes, corresponding research on ligand-protected plasmonic gold NPs such as thiolated gold nanoparticles was rather limited.[3,4,14,15,19,24–26,43,44] In recent years, thiolate-protected gold clusters have received enormous research attention as the clusters with sizes less than 2 nm seem to show interesting quantum size behavior and excitonic properties.[27,28,30,31,45–47] The advent of modern synthetic and characterization techniques has made the science of atomically precise thiolate-protected gold clusters interesting for theoreticians and experimentalists. Even thiolate protected gold clusters show plasmonic properties and $Au_{329}$ protected with hexane thiol was found to be the smallest aliphatic thiolate protected plasmonic Au

NPs.[48–50] The composition of 76.3 kDa aliphatic or aliphatic-like (SC$_2$Ph) thiolate-protected gold cluster has been found to be Au$_{329}$(SR)$_{84}$ (where SR is the thiolate ligand) by employing SC$_2$Ph, SC$_6$ and SC$_{12}$ ligands with and without Cs$^+$ adducts.[51] A closely related composition of Au$_{333}$(SR)$_{79}$ has been reported for the 76.3 kDa cluster based on Cs$^+$ adduct peaks and it remains to be addressed.[49,52] Recently, hexane thiolate protected Au$_{\sim 1400}$ and Au$_{\sim 2000}$, two new sizes in aliphatic thiolate protected plasmonic Au NPs of size ~3.6 nm and ~3.8 nm, respectively, have been reported.[53,54]

One common way to study plasmons in metal nanoparticles is via the use of ultrafast transient spectroscopy and corresponding pump-power dependence that yields information of electron dynamics.[29,43,52,55–62] Comprehensive understanding of the relaxation processes of metal nanoparticles yielded valuable information that was useful for applications like exciton-plasmon energy transfer and optical switching[56–61] Link and El-Sayed investigated pump-power, size and shape effects on electron-phonon (e-p) relaxation dynamics in gold and silver nanospheres and nanorods in size range from 10 to 100 nm.[43] The studies demonstrated that e-p relaxation dynamics was independent of the size and shape of NPs.[43] Similarly, Hodak and coworkers reported a size independent electron-phonon coupling in Au nanoparticles in the 2.5 to 8 nm range in aqueous solution.[56] Link *et al.* embedded 14.5 nm and 12.1 nm Au NPs in MgSO$_4$ powder and solution to test the effect of the surrounding medium on the recovery time of plasmon bleach.[57] They found that electron-phonon relaxation time of Au NPs in MgSO$_4$ powder is higher than in solution by a factor of 2 that was assigned to the effect of the medium. In a recent work, *e-p* dynamics of the different sized gold nanoparticles conjugated with the chemical-ligand has been studied.[52] By mapping the bleach recovery dynamics of these nanoparticles, Zhou *et. al.* demonstrated that Au clusters with diameters larger than 2.3 nm show metallic behavior while NPs with a diameter smaller than 1.7 nm display pure molecular behavior. In addition, it was shown that particles within 1.7 to 2.3 nm range show both metallic and excitonic behavior. In a recent study, the first crystal structure of plasmonic gold nanocrystal, TBBT protected Au$_{279}$ has been reported.[63] More recently, our group studied the optical properties of $Au_{36}(SR)_{24}$, $Au_{44}(SR)_{28}$, $Au_{133}(SR)_{52}$ and $Au_{279}(SR)_{84}$ using steady-state and transient absorption, time dependent density functional theory and density of state calculations.[29] By observing power dependent bleach recovery kinetics in $Au_{279}(SR)_{84}$, we reported that it is the smallest gold thiolate nanoparticle that has shown metallic behavior and support localized SPR.

The type of ligand (aliphatic/aromatic/bulky) being used determines the structure and properties of metal clusters.[64,65] Although significant research has focused on molecule-like properties of gold clusters and studied how ligands alter their optical properties, the same influence of ligands on the e-p relaxation dynamics of ultra-small clusters has not been addressed. In addition, previous theoretical studies have neglected the chemical ligand effect on the conductivity of Au NPs and consequently on their e-p relaxation. In the present work, electron-phonon relaxation dynamics of smallest sized plasmonic gold clusters conjugated with different thiolate ligands was studied using ultrafast pump-probe spectroscopy. To explore the effect of aromatic and aliphatic ligands on the electron dynamics, investigations were carried out on $Au_{279}$ (d ~ 2.1 nm) passivated with TBBT and $Au_{329}$ (d ~ 2.2 nm) passivated with phenylethane thiol ($SC_2Ph$), and hexane thiol ($SC_6$). To understand the ligand effects theoretically, the free electron density of the clusters was modeled using three-layered Mie theory[66,66], inspired by Peng et al.[41] The size effect on the electron dynamics in aliphatic thiolate protected plasmonic gold clusters was investigated by studying the optical properties of hexane thiolate protected $Au_{329}$ (d ~ 2.2 nm), $Au_{\sim 1400}$ (d ~ 3.6 nm) and $Au_{\sim 2000}$ (d ~ 3.8 nm).

## 2. Methods

### 2.1 Synthesis and Characterization of Clusters

$Au_{279}(TBBT)_{84}$, $Au_{329}(SC_2Ph)_{84}$, $Au_{329}(SC_6)_{84}$, $Au_{\sim 1400}(SC_6)$ and $Au_{\sim 2000}(SC_6)$ were synthesized following the reported protocols from ref.[63], ref.[68], ref.[50], ref.[53], and ref.[54], respectively. Citrate-stabilized gold nanoparticles were synthesized following a procedure published elsewhere.[69]

## 2.2 Transient absorption measurements

Femtosecond transient absorption measurements were carried out at the Center for Nanoscale Materials, Argonne National Laboratory.[50] Briefly, a Spectra Physics Tsunami Ti:Sapphire, 75 MHz oscillator was used to seed a 5 KHz Spectra-physics Spit-Fire Pro regenerative amplifier. 95% of the output from the amplifier is used to pump a TOPAS optical parametric amplifier, which is used to provide the pump beam in a Helios transient absorption setup (Ultrafast Systems Inc.). A pump beam of 370 nm was used for the measurements. The remaining 5% of the amplified output is focused onto a sapphire crystal to create a white light continuum that serves as the probe beam in our measurements (450 to 700 nm). The pump beam was depolarized and chopped at 2.5 kHz and both pump and probe beams were overlapped in the sample for magic angle transient measurements.

**2.3 Theoretical Modeling**

From the transient absorption measurements, it was shown that the ligand plays an important role in the surface plasmon bleach spectral width as well as electron-phonon relaxation time and coupling strength. To understand this behavior, the SPR bleach was modeled with three-layered Mie theory [41,66,67] We considered the modeling geometry as a single NP because of the fact all of the investigated clusters characterized by mass spectrometry[50,53,54,63,68] and have very narrow size distribution. In our simulation, the NP has been considered to be a sphere in order to take into account all of the possible orientation of the icosahedral NPs to the incident light beam. In addition, as these NPs are coated with the ligand layer which alters the conductivity of the surface and core region of the NP differently, the three-layered Mie theory has been used to simulate optical properties of them. As illustrated in Figure 1, the three layers structure of gold clusters consists of a core region with a diameter of $d_{core} = d_{NPs} - 2t$, skin layer with a thickness of t and chemical ligand with a width of $l$. We fixed t to be roughly the thickness of an atomic Au layer (or Au-Au bonding length), t = 0.28 nm. In addition, in our simulation, the diameter of clusters were obtained from literature and are taken as: $d_{279}$ = 2.1 nm,[63] $d_{329}$ = 2.2 nm,[68] $d_{\sim1400}$ = 3.6 nm,[53] $d_{\sim2000}$ = 3.8 nm.[54] The thickness of chemical ligands were fixed to $l = 0.8$ nm. Finally, the refractive index was used as 1.492, 1.43, and 1.375 for the TBBT, SC$_2$Ph and SC$_6$ ligands, respectively.[70]

By reducing the size of the NP (d< 40 nm), the mean free path of the conduction electrons will decrease leading to increased collision rate of the electrons with the surface of the NP as well as more damping of the plasmon band. To accurately simulate the optical properties of these NPs, modification of damping constant of the bulk is necessary with shrinking dimension of the NPs. Existing classical models have described these changes by expressing size dependent damping factor as [42,44,71,72]

$$\Gamma(r) = \Gamma_0 + \frac{AV_F}{r} \quad (1)$$

where $\Gamma_0$ represents the bulk damping constant, $V_F = 1.39 \times 10^6$ ms$^{-1}$ is Fermi velocity, r is the effective radius of the NP and A is an empirical damping factor ranging from 0.1 to 2, depending on the experimental data and physical model.[42,44,71,72]

The modified permittivity for the NPs with 10 nm < d < 40 nm can be expressed as [42]

$$\varepsilon_{Np}(\omega, r) = \varepsilon_{bulk}(\omega) + \frac{\omega_p^2}{\omega^2 + i\Gamma_0\omega} - \frac{\omega_p^2}{\omega^2 + i\Gamma(r)\omega} \quad (2)$$

where $\varepsilon_{bulk}$ is permittivity of the bulk metal, $\omega_p$ is plasma frequency ($\omega_p = \sqrt{ne^2/m\varepsilon_0}$), and $\omega$ is the angular frequency of the incident light.

When the size of NP is less than 10 nm, the quantum effect must be taken into account. For this small sized region, the energy levels of the conduction band are discretized, and only certain electronic or plasmonic transitions are allowed.[71,73] By considering these quantum effects, the permittivity of the quantum-sized NP can be written as [71,73]

$$\varepsilon_{QNP}(\omega) = \varepsilon_{IB} + \omega_p^2 \sum_i \sum_f \frac{S_{if}}{(\omega_{if}^2 - \omega^2 - i\Gamma(r)\omega)} \qquad (3)$$

where the sum is taken over all initial and final states. In Eq. 3, $\varepsilon_{IB}$ is the contribution of interband transitions that can be obtained elsewhere,[74] and $\omega_{if}$ and $S_{if}$ represent transition frequency and oscillator strength between initial and final states of electrons, respectively.

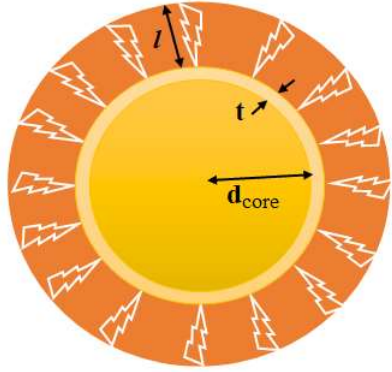

**Figure 1.** Schematic illustration of three layers structure of spherical NPs consists of the core region of diameter $d_{core}$, and skin and surrounding chemical ligand with a thickness of t and *l*, respectively.

As discussed above, conjugating the metallic NP with ligand reduces the density of the free electrons in the conduction band of the metal due to chemical bonding between the NP and the ligand.[41] These chemical interactions have the most effect on the outermost layer of the NPs by decreasing the number of free electrons. This caused a shift of bleach maximum to longer wavelength and dampened the plasmon band.[41] To consider the ligand effects on the permittivity of the NPs, it is necessary to modify the electronic conductivity of the core and shell by a factor of $g^2_{core}$ and $g^2_{skin}$, respectively. By applying these modifications to the equations (2) and (3), the permittivity of NPs can be represented as below for classical sized NPs (d ≥ 10 nm):

$$\varepsilon_{NP}(\omega, r) = \varepsilon_{bulk}(\omega) + \frac{\omega_p^2}{\omega^2 + i\Gamma_0 \omega} - \frac{g^2 \omega_p^2}{\omega^2 + i\Gamma(r)\omega} \quad (4)$$

where g is representing the plasma frequency correction factor. g is $g_{core}$ for core region and g is $g_{kin}$ for skin region.

On the other hand, the modified electric permittivity for quantum sized NPs (d ≤ 10 nm) can be expressed as:

$$\varepsilon_{QNp}(\omega) = \varepsilon_{IB} + g^2 \omega_p^2 \sum_i \sum_f \frac{S_{if}}{(\omega_{if}^2 - \omega^2 - i\Gamma(r)\omega)} \quad (5)$$

where, $\varepsilon_{IB}$ is inter-band contribution, $\omega_p$ is plasma frequency, $\omega_{if}$ is inter-level transition frequency between the initial and final state, $S_{if}$ represent the oscillator strength between initial and final states. In this work, the Eq. 4 and Eq. 5 have been applied to simulate the absorption spectrum of the $Au_{13nm}$ and all ligand-protected clusters, respectively.

## 3. Results and Discussion

### 3.1 Ultrafast transient absorption measurements

Optical absorption spectra of the investigated gold clusters are shown in Figure S1. The linear absorption maxima of smaller sized ligand-protected gold clusters is appearing around 500 to 510 nm, much below the expected surface plasmon absorption of larger sized gold nanoparticles (Figure S1). This condition is arising because of the overlapping of inter-band transitions with the surface plasmon absorption, thereby shifting the shape and maximum of absorption maximum to higher energies. Because of this reason, the linear absorption of the thiolate protected clusters is unable to represent the true surface plasmon absorption. Thus, to model the ligand effects on the SPR, we used the bleach spectrum obtained from transient absorption measurements. Transient absorption (TA) measurements were carried out after excitation at 370 nm for all clusters and probing in the visible region. Pump-power dependent transient absorption measurements were carried out to monitor the electron dynamics. Parts A, B, C and D of Figure 2 show the TA spectra at pump energy of 120 nJ for $Au_{279}$ (TBBT), $Au_{329}$ ($SC_2Ph$), $Au_{329}$ ($SC_6$) and $Au_{13\,nm}$ (Citrate), respectively. Excitation of the nanoparticles create hot electrons and their relaxation is often monitored using pump-power dependent transient absorption measurements.[43,54,55,59] As observed in Figure 2, a negative absorption centered around 544 to 550 nm with two positive wings was observed for all investigated gold clusters which is consistent with the literature reports of transient absorption of plasmonic gold nanoparticles.[52]

Similar spectral features were also observed for $Au_{\sim1400}$ ($SC_6$), $Au_{\sim2000}$ ($SC_6$), and $Au_{13\,nm}$ (Citrate) (see Figure S2, and Figure 2D). The bleach maximum is attributed to the SPR absorption of the clusters and the accurate maximum was determined by the fit of the bleach curve.[43,52,59,62,75,76] The bleach maximum has shifted to higher energies for all samples with an increase in time, except for $Au_{329}$ ($SC_2Ph$) that has shown two peaks (see Figure 2, and Figure S2). Presence of two peaks for $Au_{329}$ ($SC_2Ph$) in the bleach spectrum was also observed by Zhou and coworkers and is quite interesting to see why such behavior was observed for this cluster.[52]

The shift of the bleach maximum to the higher energies is attributed to a decreased electronic temperature of the illuminated clusters because of the heat transfer to phonon bath. From Figure 2, it can be seen that the bleach maximum of the ligand-passivated gold clusters is shifted to lower energies as well as broadened when compared to ligand-free gold nanoparticles.[41] The FWHM depends on many factors, such as the size of nanoparticles, polydispersity in size and shape and chemical bonding between the nanoparticles and ligand.[41] Although the ligand exchange can diphase the oscillations of the plasmonic electrons by adding an additional decay channel due to the transient in and out tunneling in interfacial orbitals with the NP and ligand,[77–79] the size of the NPs plays a dominant role in plasmonic damping (See Figure 3). As demonstrated in the Figure 3 and Figure S3, the FWHM of the smaller samples (Au$_{279}$ (TBBT), Au$_{329}$ (SC$_2$Ph) and Au$_{329}$ (SC$_6$)) at a delay time of 0.5 ps were determined to be ~ 92 nm, which decreases with further increasing the size of the samples and reaches to ~ 53 nm for 13 nm-Au.

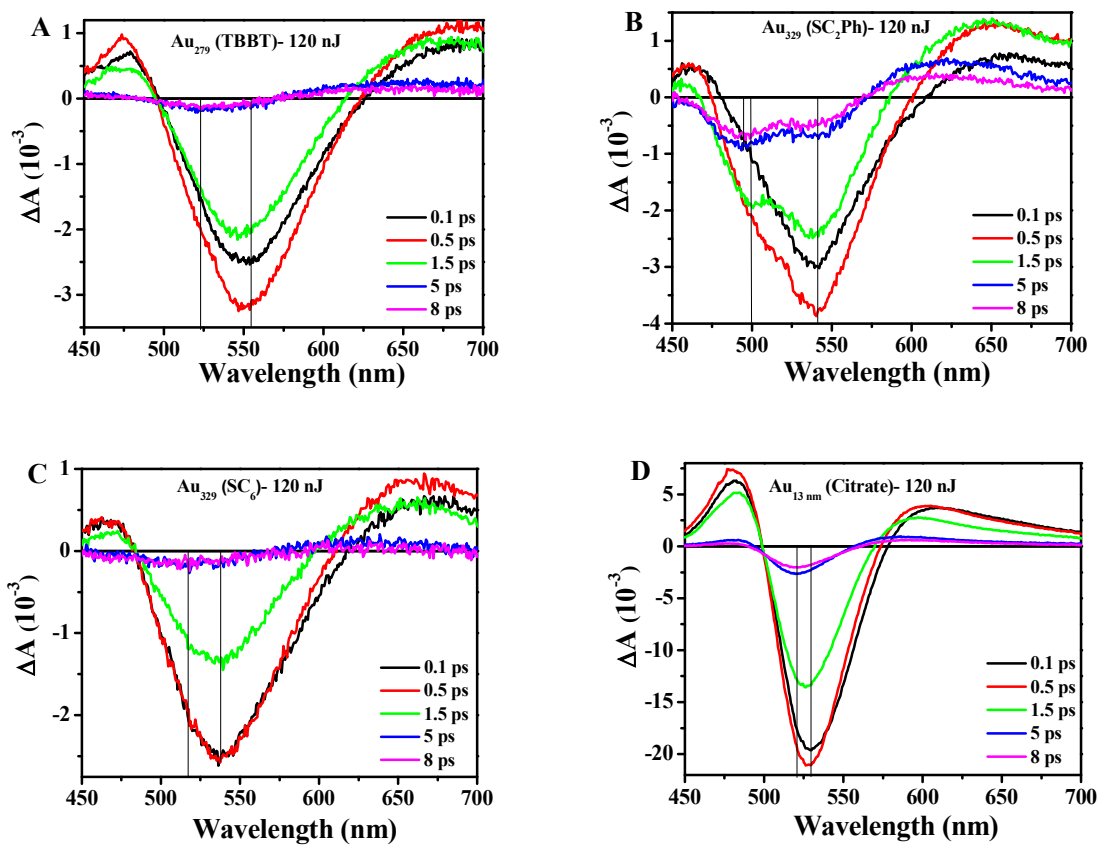

**Figure 2.** Transient absorption spectra at different time delays for (A) Au$_{279}$ (TBBT), (B) Au$_{329}$ (SC$_2$Ph) (C) Au$_{329}$ (SC$_6$) and (D) Au$_{13\,nm}$ (citrate) after excitation at 370 nm.

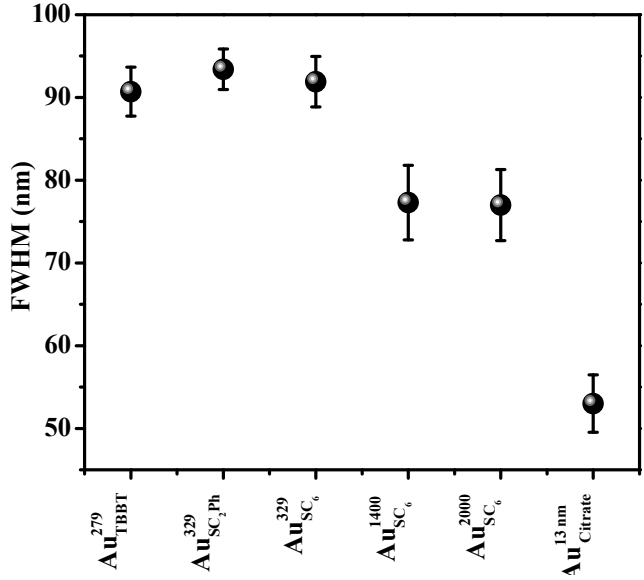

**Figure 3.** FWHM of the studied samples at a time delay of 0.5 ps for the pump energy of 120 nJ.

As the central aim of the study is to understand the influence of passivating ligand on the electron dynamics, transient bleach recovery dynamics was monitored as a function of pump power. Pump-power dependence of bleach recovery kinetics is a key feature for plasmonic nanoparticles as it alters the electronic temperature of the nanoparticles. Figure S4 A-F shows the bleach recovery kinetic traces for the investigated clusters at different pump powers. As expected with an increase in pump-power, the bleach signal increases as the local temperature of cluster increases. Increasing pump energy will heat up more electrons and decelerating the e-p coupling leads to slowing down of electron relaxation.[52,80] The pump-power dependence of bleach recovery can be modeled using a two-temperature model:[81–84]

$$C_e(T_e)\frac{\partial T_e}{\partial t} = -\gamma(T_e - T_l) + \alpha N \qquad (6)$$

$$C_l\frac{\partial T_e}{\partial t} = \gamma(T_e - T_l) + \beta N \qquad (7)$$

where $C_e, T_e$ and $C_l, T_l$ are heat capacity and temperature of the electron gas and lattice, respectively. $\alpha N$ describes the heating of the electron gas by the initial nonthermalized electrons, $\beta N$ represents the direct coupling between the non-thermalized electrons and the lattice occurs during the electron gas thermalization process,[84] and γ is electron-phonon coupling constant.[56,85–87]

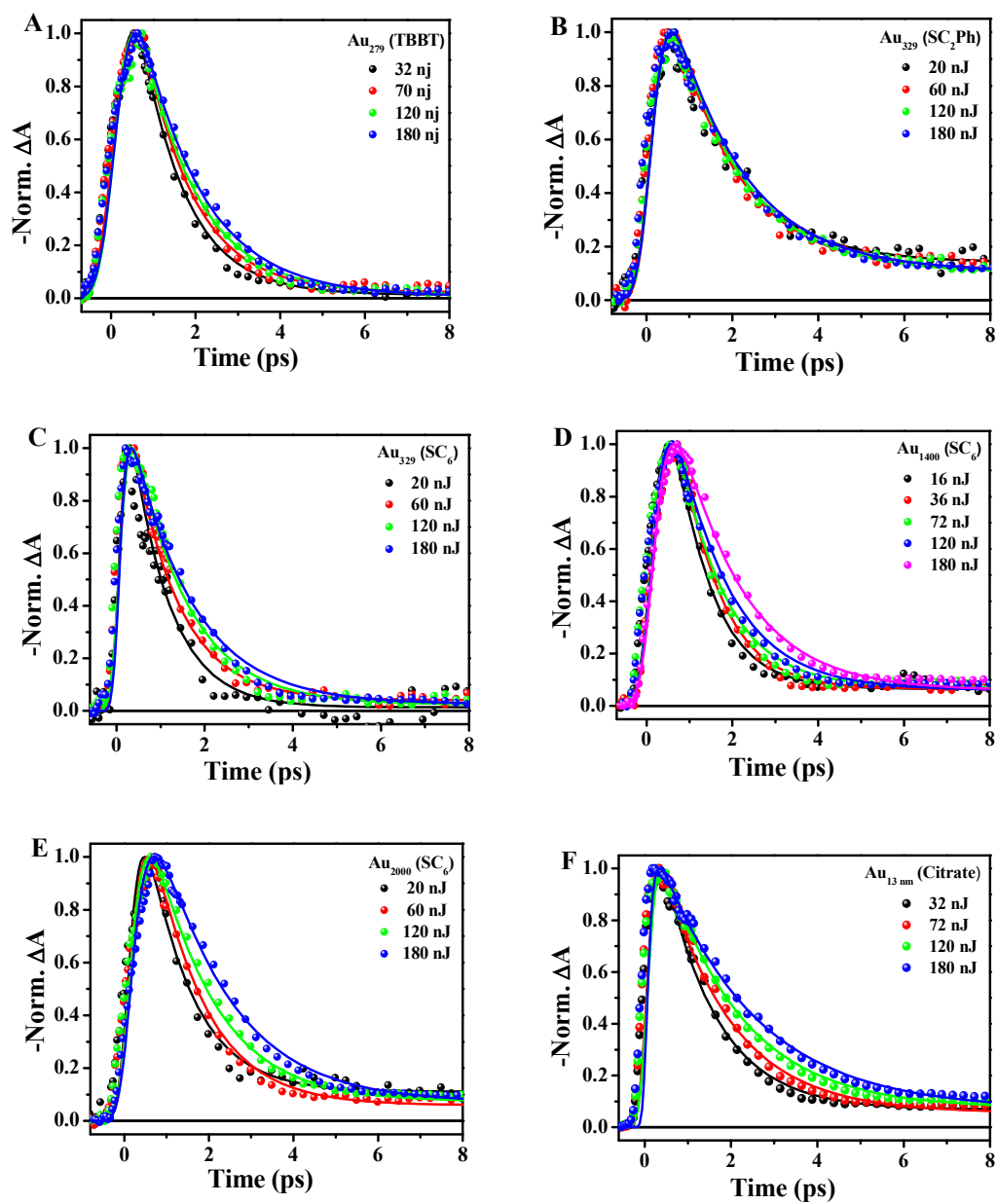

**Figure 4.** Normalized bleach recovery kinetics (-ΔA) of (A) $Au_{279}$ (TBBT), (B) $Au_{329}$ ($SC_2Ph$), (C) $Au_{329}$ ($SC_6$), (D) $Au_{\sim1400}$ ($SC_6$), (E) $Au_{\sim2000}$ ($SC_6$) and (F) $Au_{13\,nm}$ (Citrate) at different pump powers.

The electron heat capacity of the metals is proportional to the electron gas temperature. Increasing the temperature of electron gas reduces electron-phonon effective coupling rate ($\frac{\gamma}{C_e(T_e)}$) and deaccelerating electron-phonon coupling time. The normalized bleach recovery (-Norm. ΔA) for the investigated clusters is shown in Figure 4 A-F. To extract electron-phonon relaxation for each cluster, the bleach recovery dynamics was fitted (see Figure 4) using:[52,88,89]

$$\Delta A(t) = \int_{-\infty}^{\infty} H(\tau) \left[ A(1 - e^{-\tau/\tau_{ee}})e^{-\tau/\tau_{ep}} + B\left(1 - e^{-\tau/\tau_{ep}}\right) \right] e^{-(\tau-t)^2/\tau_0^2} d\tau \qquad (8)$$

where $H(\tau)$ is Heaviside function, A and B are e-e and e-p scattering amplitudes, where $|B| < |A|/10$. $\tau_{ee}$ is e-e coupling time and is in the order of few hundred femtoseconds, and $\tau_{ep}$ is e-p coupling time and is in order of ~ 1 ps. $\tau_0$ is instrument response obtained by the cross correlation of the pump and probe beams. From the intercept and slope of the linear fit of e-p relaxation lifetime versus pump energy, intrinsic e-p coupling time and e-p coupling strength for all samples were determined and provided in Table 1.

Figure 5A shows the plot of calculated $\tau_{e-p}$ at different pump energies for $Au_{279}$ (TBBT), $Au_{329}$ ($SC_2Ph$), $Au_{329}$ ($SC_6$) and $Au_{13\,nm}$ (Citrate) nanoparticles. As seen from the figure, the slopes and intercepts varied for different clusters. $Au_{279}$ (TBBT) and $Au_{329}$ ($SC_6$) has shown an intercept of close to 1 ps and 0.8 ps, respectively, while $Au_{329}$ ($SC_2Ph$) has shown an intercept of 1.4 ps. Similar higher intercept was also observed by Zhou et al[52] for $Au_{329}$ ($SC_2Ph$). One another interesting observation is the difference in slopes for $Au_{279}$ (TBBT), $Au_{329}$ ($SC_6$) and 13 nm Au cluster. The slope is smaller for $Au_{279}$ (TBBT) when compared to $Au_{329}$ ($SC_6$), which is much smaller than that of 13 nm Au nanoparticle (See Table 1). This is attributed to differences in electric conductivities because of the nature of passivating ligands. Aromatic passivating ligands seem to have smaller slopes and higher intercepts when compared to hexane thiol ligand. Also, the plot of e-p relaxation as a function of pump power for different sized SC$_6$ protected gold cluster is shown in Figure 5B. It can be observed from the Figure 5B and Table 1 that with an increase in size, total electric conductivity increases slightly due to a decrease of the surface to volume ratio, and thereby increasing the intercept (intrinsic electron- phonon coupling time) and slope (electron-phonon coupling strength). However, the slope is definitely smaller when compared to Au 13 nm nanoparticle that was passivated with a surfactant, again signifying the importance of the ligand on electron-phonon relaxation dynamics.

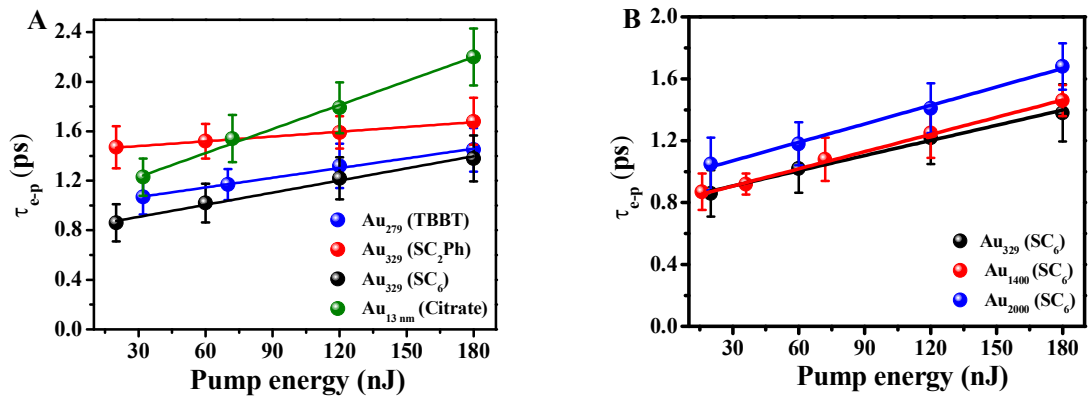

**Figure 5.** Electron- phonon relaxation as a function of pump pulse energy for (A) Au$_{279}$ (TBBT), Au$_{329}$ (SC$_2$Ph), Au$_{329}$ (SC$_6$) and Au$_{13\,nm}$ (Citrate), (B) Au$_{329}$ (SC$_6$), Au$_{\sim1400}$ (SC$_6$) and Au$_{\sim2000}$ (SC$_6$).

**Table 1.** Electron-phonon coupling time and coupling strength obtained from power-dependent bleach recovery kinetics of the investigated clusters

|  | Au$_{279}$- TBBT | Au$_{329}$- SC$_2$Ph | Au$_{329}$- SC$_6$ | Au$_{\sim1400}$- SC$_6$ | Au$_{\sim2000}$- SC$_6$ | Au$_{13\,nm}$ -Citrate |
|---|---|---|---|---|---|---|
| **Slope** | $(2.6 \pm 0.3) \times 10^{-3}$ | $(1.2 \pm 0.2) \times 10^{-3}$ | $(3.2 \pm 0.3) \times 10^{-3}$ | $(3.7 \pm 0.3) \times 10^{-3}$ | $(3.9 \pm 0.3) \times 10^{-3}$ | $(6.0 \pm 0.4) \times 10^{-3}$ |
| **Intercept** | $1.0 \pm 0.2$ | $1.4 \pm 0.2$ | $0.8 \pm 0.2$ | $0.8 \pm 0.1$ | $1.0 \pm 0.2$ | $1.0 \pm 0.2$ |

Among all investigated clusters, unique behavior was observed for Au$_{329}$ (SC$_2$Ph). The transient bleach of Au$_{329}$ (SC$_2$Ph) has shown two negative bleach peaks around ~ 493 and 540 nm. A similar result was reported by Zhou and coworkers and assigned it to excitonic and plasmonic behavior of the cluster.[52] Interestingly, a closer look to the transient absorption spectrum of the samples in Figure 2 reveals that transient bleach of the Au$_{329}$ (SC$_2$Ph) is higher and wider than Au$_{279}$ (TBBT) and Au$_{329}$ (SC$_6$), respectively. This unusual result can be ascribed to specific Au-π interaction between aromatic ring in phenylethane thiol and surface gold atom of Au$_{329}$ (SC$_2$Ph), significantly reducing the electric conductivity of the surface and core layers of the cluster (see Figure 6). This specific Au-π interaction potentially reduces the free electron density of Au$_{329}$ significantly (refer to theoretical modeling section) and thereby creating a hybrid state[52] (metallic- molecular) for it. Additional modeling is necessary to understand the influence of passivating ligands on electron density in these clusters.

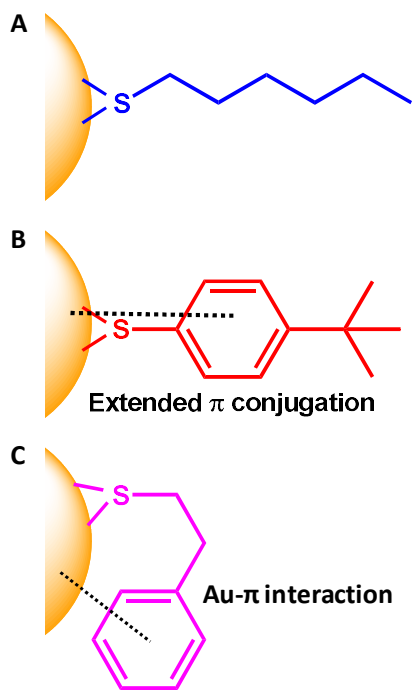

**Figure 6.** Cartoon diagram depicting the interaction of (A) SC$_6$, (B) STBBT, and (C) SC$_2$Ph, with the surface of gold atoms.

As the transient bleach at initial times represents the SPR bleach, spectrum at a time delay of 500 fs has been used into the fit to the theoretical absorption spectra (See Figure S5) in order to obtain the electronic conductivity correction factors and corresponding $g_{core}$ and $g_{skin}$.[43,52,59,62,75,76] Obtained $g_{core}^2$ and $g_{skin}^2$ from the analysis for different samples are shown in Figure 7A. A closer inspection of Figure 7A reveals interesting trends. Both $g^2_{core}$ and $g^2_{skin}$ values for aromatic passivating ligands are much smaller when compared to hexane thiol passivating ligand. Also, the $g^2_{skin}$ obtained for $SC_2Ph$ is much smaller than TBBT, and the results are consistent with experimentally observed electron dynamics in these clusters. The modeling suggests that it is the electron density difference, which is crucial for electron dynamics in these clusters. Due to the direct interaction between the chemical ligand and surface layer, electron conductivity of the outermost layer of the metallic NPs is decreased more when compared to the core region. In agreement with our experimental results, aromatic ligands, especially $SC_2Ph$, lowered the conductivity of the NPs more when compare to hexane thiol. As mentioned above, owing to the specific Au-π interaction between the surface gold and $SC_2Ph$ ligand, the free electron density of the skin and core regions of $Au_{329}$ are lowered by a factor of 0.64 and 0.9, respectively. This significant reduction of the electron conductivity, especially for the outermost layer, is a further justification of deaccelerated electron-phonon dynamics in $Au_{329}$ ($SC_2Ph$) and is the reason for observed smaller power dependence for this cluster. Also smaller electron conductivity was also observed for $Au_{279}$ (TBBT) that can be attributed to Au-π interaction. But this interaction is probably the reason for observing plasmon behavior for this cluster even though its size is smaller when compared to $Au_{329}$.

For the clusters with same $SC_6$ passivating ligand, the $g_{skin}$ values have slightly increased with an increase in cluster size while $g_{core}$ values are the same as that of 13 nm Au NP and increased electron conductivity with increase in the size of the cluster (see Table 2). Present results show zero reduction on the number of free electrons in citrate-capped 13 nm Au nanoparticles. The total conductivity correction factor of each cluster has been calculated by:

$$g_t^2 = g_{core}^2 f_{core} + g_{skin}^2 f_{skin} \qquad (9)$$

where $f_{core}$ and $f_{skin}$ are core and skin volume fraction, respectively. As shown in Figure 7B, the total conductivity correction factor of $Au_{329}$ ($SC_2Ph$) is smaller than other samples. In addition, for the clusters with the same $SC_6$, owing to the decreasing surface to volume ratio the total modified conductivity increases with increasing size of the samples. The three-layered Mie theory results were able to accurately model the influence of ligand on electron density and their effect on plasmon quality factor. This ligand influence can have consequences for electric field enhancement offered by ligand-protected plasmonic clusters.

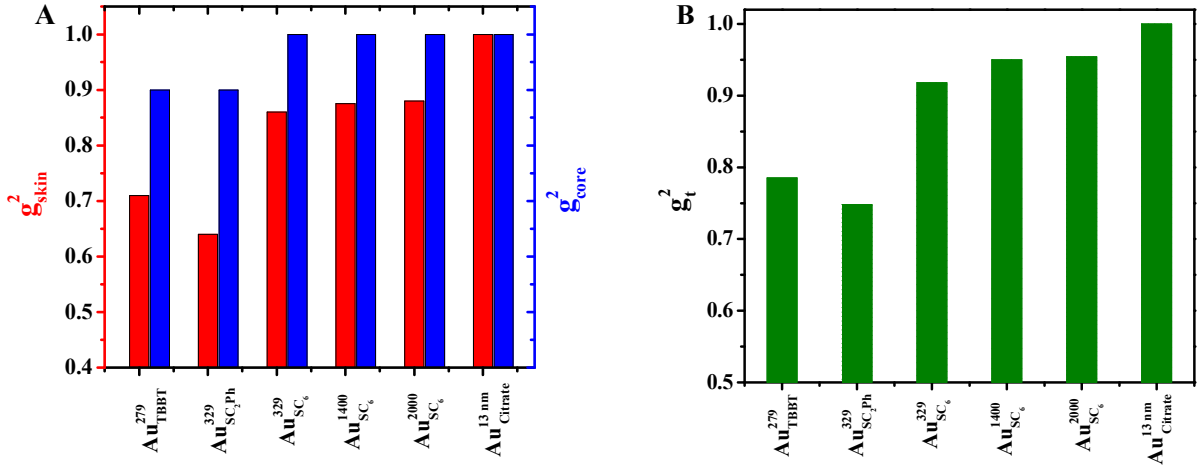

**Figure 7.** (A) Free electron density correction factor of the outermost layer ($g_{skin}^2$) and core region ($g_{core}^2$), (B) total free electron density correction factor ($g_t^2$) of the investigated Au clusters.

**Table 2**. Summary of the studied samples and their characteristics. FWHM of each sample has been at t ≈ 0.5 ps for the pump energy of 120 nJ.

| Samples | Size (nm) | FWHM (nm) | $g^2_{core}$ | $g^2_{skin}$ | $t_{e\text{-}ph}$ |
|---|---|---|---|---|---|
| **Au$_{279}$ (TBBT)** | 2.1 | 91 ± 3 | 0.9 | 0.71 | 1.0 ± 0.2 |
| **Au$_{329}$ (SC$_2$Ph)** | 2.2 | 93 ± 3 | 0.9 | 0.64 | 1.4 ± 0.2 |
| **Au$_{329}$ (SC$_6$)** | 2.2 | 92 ± 3 | 1 | 0.86 | 0.8 ± 0.2 |
| **Au$_{\sim1400}$ (SC$_6$)** | 3.6 | 77 ± 5 | 1 | 0.875 | 0.8 ± 0.1 |
| **Au$_{\sim2000}$ (SC$_6$)** | 3.8 | 77 ± 4 | 1 | 0.88 | 1.0 ± 0.2 |
| **Au$_{13\,nm}$ (Citrate)** | 13 | 53 ± 4 | 1 | 1 | 1.0 ± 0.2 |

## 4. Conclusions

The electron-phonon relaxation dynamics measurements carried out on smallest sized plasmonic gold clusters have shown interesting trends with regards to the effect of the ligand on the electron dynamics. The results show that the gold clusters with aromatic passivating ligands (TBBT and SC$_2$Ph) have higher plasmon bleach and smaller electron-phonon coupling strength when compared to the gold cluster with hexane thiol passivating ligand. The electron dynamics of gold clusters with hexane thiol are in line with the results obtained for citrate stabilized gold clusters as well as larger gold clusters with the same passivating ligand. The different electron-dynamics observed for aromatic passivating ligands is ascribed to reduced electron conductivity offered by Au-π interaction and conjugation. The electron conductivity was modeled with three-layered Mie theory and the results have shown that all ligand passivated gold clusters have smaller surface conductivity when compared to citrate stabilized gold nanoparticles. Within the ligand passivated clusters, TBBT, and SC$_2$Ph ligand passivated clusters have much smaller electron conductivity that can be attributed to the way the aromatic ligands interact with surface gold atoms. Especially, unique results were obtained for Au$_{329}$ (SC$_2$Ph) that has shown smaller surface electron conductivity as well as two bleach maxima in transient absorption spectra and can be attributed to specific π interaction between the ligand and Au on the surface creating a hybrid molecule/metallic state. However, similar π-conjugation was probably the reason for observing plasmon behavior for Au$_{279}$, which is smaller in size when compared to Au$_{329}$. In addition, it was found that the electron-phonon coupling of the samples conjugated with same chemical ligands (Au$_{329}$ (SC$_6$), Au$_{\sim1400}$ (SC$_6$) and Au$_{\sim2000}$ (SC$_6$)) depend on the size of the cluster wherein the electron conductivity increased with increase in the size of the cluster.

**5. Supporting Information Available**. Syntheses and characterizations of gold clusters, additional transient absorption spectral results are provided.

AUTHOR INFORMATION

Corresponding Author

Email: rama.guda@wmich.edu; dongil@yonsei.ac.kr; amal@olemiss.edu

6. ACKNOWLEDGMENT

M. S. thanks physics department of Western Michigan University for financial support. G.R. acknowledges the support of Western Michigan University-FRACAA. G. R. acknowledges Dr. Gary Wiederrecht, Argonne national laboratory for help with transient absorption measurements. Use of the Center for Nanoscale Materials, an Office of Science user facility, was supported by the U.S. Department of Energy, Office of science, Office of Basic Energy Sciences, under contract no: DE-AC02-06CH11357. D.L. acknowledges support by the Korea CCS R&D Center (KCRC) Grant (NRF-2014M1A8A1074219) and NRF Grants NRF-2017R1A2B3006651 and NRF-2018M3D1A1089380. NSF-CHE-1808138 and NSF-CHE-1255519 supported the work performed by N.S., C.K., and A. D.

# TOC Graphic

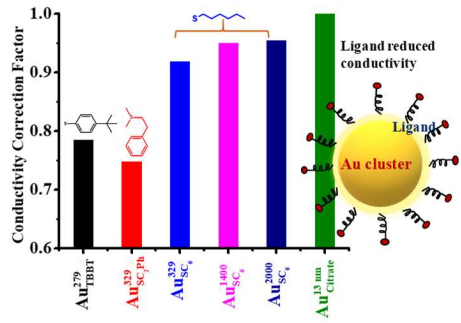